\pdfoutput=1
\documentclass{JINST}
\usepackage{aas_macros}
\usepackage{amssymb,amsmath}
\usepackage{mathtools}
\usepackage{algorithmic}
\usepackage{algorithm}
\usepackage{bm,bbm}
\usepackage{epstopdf}
\usepackage{graphicx}
\usepackage{placeins}

\DeclareMathOperator{\prox}{prox}

\title{Sparse representations and convex optimization as tools for LOFAR radio interferometric imaging}

\author{Julien N. Girard$^a$\thanks{Corresponding author.}, Hugh Garsden$^b$, Jean Luc Starck$^a$, St\'ephane Corbel$^a$, Arnaud Woiselle$^c$, Cyril Tasse$^d$, John P. McKean$^e$ and J\' er\^ome Bobin$^a$

\\
\llap{$^a$}CEA Saclay, Service d'Astrophysique, B\^at. 709, Orme des Merisiers, 91190, Gif-Sur-Yvette, France, E-mail: \email{julien.girard@cea.fr}\\
\llap{$^b$}Harvard-Smithsonian Center for Astrophysics, 60 Garden Street, Cambridge, MA 02138, USA\\
\llap{$^c$}Sagem (Safran), 27 rue Leblanc, 75512 Paris Cedex 15, France\\
\llap{$^d$}GEPI, Observatoire de Paris, CNRS, Universit\'e Paris Diderot, 5 place Jules Janssen, 92190 Meudon, France\\
\llap{$^e$}Netherlands Institute for Radio Astronomy (ASTRON), Postbus 2, 7990 AA Dwingeloo, The Netherlands \\
E-mail: \href{mailto:julien.girard@cea.fr}{julien.girard@cea.fr}
  }

\abstract{Compressed sensing theory is slowly making its way to solve more and more astronomical inverse problems. We address here the application of sparse representations, convex optimization and proximal theory to radio interferometric imaging. First, we expose the theory behind interferometric imaging, sparse representations and convex optimization, and second, we illustrate their application with numerical tests with SASIR, an implementation of the FISTA, a Forward-Backward splitting algorithm hosted in a LOFAR imager. Various tests have been conducted in Garsden et al., 2015. The main results are: i) an improved angular resolution (super resolution of a factor $\approx$ 2) with point sources as compared to CLEAN on the same data, ii) correct photometry measurements on a field of point sources at high dynamic range and iii) the imaging of extended sources with improved fidelity. SASIR provides better reconstructions (five time less residuals) of the extended emission as compared to CLEAN. With the advent of large radiotelescopes, there is scope for improving classical imaging methods with convex optimization methods combined with sparse representations.}


\begin{document}


\section{Introduction}

The emergence of the new sampling theorem known as \textit{Compressed Sensing} \cite{candes06,donoho06} enabled to build, along with convex optimization and sparse representations, a solid framework to solve a variety of inverse problems (such as imaging) which are daily found in medical imaging, biology and recently in astronomy (e.g. \cite{bobin}). In this framework, new methods appeared for robust and fast image reconstruction which are crucial when using giant radio interferometers such as LOFAR and SKA.

We present here a summary of the basics in radio interferometry in Section \ref{radiointerferometry}, from the measurement with an antenna (\S\ref{radiomeasure}) to the method used to produce radio maps (\S\ref{aperturesynthesis}). In Section \ref{sparse}, we introduce the notion of sparse representation (\S\ref{sparserep}) and one current method to solve convex optimization problems (\S\ref{inverseproblem}). In Section \ref{application}, we formulate the imaging problem as an inverse problem in the CS framework (\S\ref{recasting}), and we illustrate the principle of reconstruction with some numerical tests made in the scope of radio imaging with modern interferometers (\S\ref{numericaltests}).

\section{Principles of radio interferometry}
\label{radiointerferometry}
	\subsection{From a single dish to antenna arrays}
	\label{radiomeasure}
		\subsubsection{Limitation of single dish observations}
		A radio antenna is an electromagnetic transducer which converts electromagnetic waves (and often keeping the information about polarization) into current/voltages signal which then can be acquired, processed and sampled. It can be simply represented by a thermometer which thermalizes with a brightness temperature over a certain region of the sky, the antenna Field Of View (FoV). Depending on its geometry and on the observing wavelength, each antenna has a directional gain pattern (or beam) which describes how the antenna response varies across the FoV, away from the pointing center. This pattern also varies with time as well as with the electrical conditions of the antenna surroundings. Classical (directive) radio antennas have a paraboloid reflector to condense the radio signal collected over an effective area\footnote{which possibly differs from the physical collecting area} ($A_\text{eff}$) in an unique focal point. The beam pattern can be decomposed in the main lobe, which defines the angular resolution of the observation, and side lobes, in which radio frequency interferences (RFI) can alter the astrophysical signal. Because there is a link between the measured Point Spread Function (PSF) of the telescope and the aperture diameter $D_{dish}$ (through the Wiener-Khintchine theorem), a large collecting area brings a higher angular resolution $\delta\theta$ on the sky following the well-known relation $\delta\theta \propto \frac{\lambda}{D_{dish}}$ .
		
These directive antennas are mechanically steerable and their mechanical operability and resistance are the main factors which limit their maximum size. The Green Bank Telescope (GBT) and the Radio Telescope Effelsberg (RTE) are the largest steerable dishes (of $\approx$ 100m in diameter) that can be built with metallic structure. To achieve higher angular resolution and a better sensitivity, the size of the reflector must be increased, and a straightforward way to build such instrument is to rely on natural formations (e.g. Arecibo Observatory) or to divide them in manageable pieces (e.g. Nan\c cay Radio Telescope). In the case of the Arecibo radiotelescope, different pointing directions are achievable by moving the focus point itself over the surface of the reflector(s).

		\subsubsection{Antenna arrays}
The construction of huge instruments as depicted above is itself limited by construction constraints and maintenance costs. Consequently, combining antennas of manageable size was proposed very early \cite{kraus} as a way to increase both angular resolution and sensitivity. Using radio antennas in arrays is not new, and was used since the early age of radio astronomy at very low frequencies by Karl Jansky\footnote{from which the flux density unit used in radio astronomy is defined:  1 Jansky = 1 Jy = $10^{-26}$ W.m$^{-2}$.Hz$^{-1}$} and Grote Reber who pioneered the observation of the first astrophysical radio sources.

The principle of antenna arrays is to extrapolate the Wiener-Khintchine theorem to sparsely filled apertures. It is a way to pave a larger artificial aperture  to inherit the properties of a larger radio telescope that can not be built in a single piece. By putting telescopes at different mutual distances and at a maximum distance $D_{\text{max}}$, one can mimic the effect of observing with a single dish telescope of aperture $D_{\text{max}}$. A maximum angular resolution $\delta \theta_{\text{max}} \propto \frac{\lambda}{D_{\text{max}}}$ can be achieved but the sensitivity scales only with the sum of the effective areas of the different telescopes rather than with the full area covered by the synthetic aperture.

There are two main ways to use radio antennas in an array:
\begin{itemize}
\item  in a \textbf{phased array}: all antenna signals are summed together after compensating for the relative geometric delays created between the antennas by observing in a particular direction. The incoming EM signal reaches the antenna at various times and will be temporally ``diluted'' if no phasing is performed. The summed signal results in an electronic beam pointed in the phasing direction. For an array of directive antennas (e.g. dishes), the phasing direction must match the pointing direction commanded to each individual antenna. For an array of undirective, wide FoV, antennas (e.g. dipole over the ground), an electronic, synthetic and directive beam is formed and pointed toward the direction of interest (i.e. the direction of phasing).
The phasing stage can be done chromatically by inserting phase differences $\Delta \phi_i^{t,\nu}$ to the signal of antenna i, (at time $t$ and frequency $\nu$) or achromatically using true time delays $\Delta \tau_i^{t}$ (e.g. coax cable lengths, delay lines). This resulting ``phased'' sum of the antenna signals improves the sensitivity by a factor of 1/$N_{ant}$ and the electronic beam becomes the PSF of the array.

\item in an \textbf{interferometer}: after performing the same delay compensation on each antenna signal, the signals are correlated against each other, averaged in time and frequency and stored for each baseline $(i,j)$ formed by antenna $i$ and $j$. Each baseline produces a fringe pattern multiplied by the antenna beam pattern (asuming that antennas $i$ and $j$ have identical electrical properties, which is not true in reality) to form the two-antenna interferometer PSF. After combination with all other baselines, the PSF of the whole instrument can be derived and corresponds to the PSF of the equivalent phased array.

\end{itemize}
There is a continuity between single dish radio telescopes and the filling of an artificial aperture with small aperture telescopes. Indeed, one can interpret a single paraboloid dish as a continuous phased array or interferometer of antennas \cite{perleybook} which are distributed along a parabola. This shape possesses only one focal point and acts as a natural phasing system for signals coming from the direction $\theta$ and reflecting on the different parts of the reflector. Any carved hole in the parabolic aperture will affect the distribution of sidelobes in the PSF but will not affect the maximum achievable angular resolution (in first approximation). By continuity, one can see an array (equipped with a proper phasing system) as a huge aperture mainly filled with holes but providing a similar angular resolution. This concept is called aperture synthesis, and radio interferometry relies on this principle to make imaging possible. The distribution of the antennas in the array is therefore important as it has a direct effect on the shape of the instrumental PSF. 
In practice, various effects alter the effective PSF of the array (effect of other antenna, spill-over, efficiency, mutual coupling, mechanical stability, presence of the ground, reflection, diffraction effects, ...).
Examples of interferometers are the Very Large Array (VLA, NRAO), the Westerbork Synthesis Radio Telescope (WSRT), the Low Frequency Array (LOFAR) \cite{LOFAR}, the Long Wavelength Array (LWA) \cite{lwa} and in the future, the Square Kilometre Array (SKA) \cite{ska09}.
Example of phased arrays are the Nan\c cay Decametre Array (NDA)\footnote{http://www.obs-nancay.fr/-Le-reseau-decametrique-.html}, any antenna field of a LOFAR station, the Murchison Widefield Array tiles (MWA) \cite{mwa} and the incoming NenuFAR array \cite{nenufar2012,GirardCR_2012}.
LOFAR is an instrument which couples the two concepts of arrays\footnote{For more information on LOFAR, see www.lofar.org}
: at the station level, antenna signals are combined in phased arrays, at the interferometer level, signal from stations can either be summed as a giant phased array, or correlated as a giant interferometer. 
		
	\subsection{Aperture synthesis with interferometry}
	\label{aperturesynthesis}
		\subsubsection{What is measured by an interferometer?}
		An interferometer is defined as a collection of antenna baselines formed with theoretically identical antennas.
		Each baseline $b_{i,j}$ will produce sets of correlation measurements at different times $t$ and frequencies $\nu$. Correlation between two antennas signal can be seen as the Young's holes interference experiment in optics. The Wiener-Khintchine theorem links the beam pattern to the Fourier transform (FT) of the aperture autocorrelation and is identical to the fringe pattern created by two coherent and in-phase point sources. In the scope of a planar array of two antennas, the beam pattern is a linear fringe pattern projected on the sky with geometry depending on the time, frequency, and the perpendicular projection of the baseline w.r.t. the direction of interest. An ideal interferometer samples the sky in the Fourier domain \cite{wilson09}.
		 As a classical antenna or antenna array samples the sky brightness temperature, an interferometer will sample the FT of the sky brightness through the measurement of spatial frequencies. One baseline (projected against the normal of the pointing direction) gives access to one spatial frequency of the sky brightness $B$. 
		Short baselines sample small spatial frequencies associated with large scale structures in the sky. Conversely, large baselines provide information on the small scale structures in the sky. The maximum baseline limits the maximum angular resolution of the observation at $\lambda$ with $\delta \theta_{\textrm{max}} \propto \frac{\lambda}{B_{\textrm{max}}}$.

		In a simplified framework, for a coplanar array in a small field approximation, the measured quantity between antennas (i,j) is given by Eq. \ref{eq:visibility}.
		\begin{equation}
		\underline{V_{i,j}}=\int_{FoV} B(\vec{r}) \exp(-2 i \pi \vec{b_{i,j}}.\vec{r} / \lambda) d\Omega
		\label{eq:visibility}
		\end{equation}
		
		$V_{i,j}$ is the complex (fringe) visibility, $d\Omega$ the solid angle element, $B(\vec{r})$ is the sky brightness centered around the phase center direction $\vec{r}$ (defining a projected 2D plane with variables $l$, $m$ and $n$ the direction cosines taking their origin at the pointing center), $\exp(-2 \pi \vec{b}_{i,j} .\vec{r} / \lambda)$ can be expressed analytically using the coordinates (u,v,w) and (l,m,n) which are Fourier pairs through the simple FT in Eq. \ref{eq:visibility}.
		
		By neglecting the third component $w$ in Eq. \ref{eq:visibility} in the small field approximation, each spatial frequency can be represented as a unique point in the ``u-v'' plane. This plane is then the FT of the sky. The more samples we have in the u-v plane, the more complete the knowledge of the sky FT. The limited number of antennas limits the number of measured spatial frequencies at a given time and frequency. More visibility data can be obtained by observing in various frequency bands and at various times (e.g. Earth Rotation Synthesis).
		Even with a perfect data calibration, the limited amount of measurements distorts the image of the sky, making the imaging problem an inverse problem. 
		Eq. \ref{eq:visibility} can be also seen as the integral of the product of the sky brightness $B$ (over the antenna FoV) by a fringe pattern projected on the sky, as if the sky was seen through a venetian blind. An interferometer is a set of spatial filters which measures the coefficient of the sky in a 2D Fourier basis of functions.

In reality, the framework is much more complex and the data modeling as well as the real data calibration is performed using the Radio Interferometer Measurement Equation (RIME) \cite{hamaker96p1,hamaker96p2,smirnov2011} which will not be detailed here (see \cite{garsden2015} and references therein). This framework is up to now the most accurate and most general linear framework to model how a polarized radio signal is transformed from its source to its measure in the form of visibilities. All effects are modeled by \textit{Jones} matrices which serve to express direction-independent effects (electronic gains, array phase, clock shift and drifts, etc.) and direction-dependent effects (antenna beam pattern, ionosphere, polarization, etc.). Direction-(in)dependent calibration of radio data is an active field of research in the scope of modern radio interferometry.

Given a discrete and discontinuous set of visibilities, which are the true measurements given by an interferometer, the task is to inverse the problem to recover the full original FT of $B$ and obtain $B$ through inversion.
		
		\subsubsection{From visibilities to images}
		
		As seen above, the imaging problem can be formulated as the inverse problem of Eq. \ref{eq:imaging}.
		\begin{equation}
		V= M F B + N
		\label{eq:imaging}
		\end{equation}
		
		With $V$ the measured complex visibilities, $B$ the brightness distribution of the sky, $F$ the Fourier operator to convert the sky brightness to its FT, $M$ a masking operator which accounts for the unsampled regions of the FT in the data and $N$ a generic additive noise which comprises all sources of noise. We assume here a perfectly calibrated set of visibilities $V$ which constitutes our data.
		Knowing $M$ and $F$, the imaging problem is to deduce $B$ from data $V$. $M$ is nothing more than the FT of the PSF and is often combined to a weighting function in the u-v plane. The latter enables some apodization of different visibilities depending on their noise and density, with the effect of modifying the instrumental PSF.
		In order to produce an image by an inverse 2D FT, the visibilities are first gridded using convolution functions on a regular grid. A raw inverse FT of the gridded visibilities will result in a highly distorted image, called the \textit{dirty} image, which is the convolution of the sky brightness with the instrumental PSF, the \textit{dirty} beam. It is very unlikely that any relevant scientific information could be retrieved from the \textit{dirty} image.
		It is necessary to perform a set of additional operations to produce a scientifically usable image.
Consequently, the problem reduces to a deconvolution problem with the goal to remove the effect of the instrumental response from the data. 
			
		\subsubsection{Deconvolution with CLEAN}
	
The CLEAN algorithm \cite{hogbom74} is a simple and efficient deconvolution algorithm which was used for $\approx$40 years by the radio community. 
It proposed an iterative approach which performs source detection and fractional PSF subtraction on the \textit{dirty} image.
At each iteration, it searches for the maximum peak location and subtracts a fraction (generally $\approx$10\%) of the PSF scaled to the detected peak maximum. It jointly fills a map with the detected peaks, known as \textit{CLEAN components}, whichs leads at the end to the \textit{model} map containing detected sources in the dirty image. The algorithm stops when a threshold is reached or when the maximum number of iterations has been performed.
The \textit{dirty} beam main maxima is then fitted with a 2D elliptical gaussian which serves as a regularization function, the CLEAN beam. The final CLEAN component map, the \textit{model} image, is convolved by this CLEAN beam and added to the \textit{residual} map to form the \textit{restored} image.
As the \textit{model} image is usually not directly usable for science, it is convolved with the CLEAN beam which accounts for the actual angular resolution accessible with the observation. In addition, the \textit{residual} map is combined to the results to restore potentially missed background emission (e.g. diffuse extended emission).

Many improvements of the CLEAN algorithm were developed \cite{wilnernrao} to enhance the original version in terms of computational effort, accuracy, and fidelity w.r.t. the source characteristics (e.g. Multiscale CLEAN \cite{Cornwell:msclean2008}). They generally improve the quality of the \textit{restored} images by building better models of the sky and images with lower residuals.
One of the latest version of CLEAN-based algorithms is the Multi-Scale Multi-Frequency Synthesis (MS-MFS) \cite{rau11} algorithm which accounts for the extended features of the source as well as its dependence in frequency. Other deconvolution algorithms were also developed, and are based on statistical approaches \cite{junklewitz13,sutter14}. We will focus in the following, on the methods based on sparsity.

		\subsubsection{Limitations of classical imaging}
With modern interferometers such as LOFAR and SKA, simplifying hypotheses do not hold any longer such as the small field and the coplanar array approximations \cite{cornwell1992}. At low frequencies (<300 MHz), the antenna beam pattern widens and becomes sensitive to a larger number of radio sources which will impact the visibilities. Correct calibration requires an advanced knowledge of the sky and often necessitates wide-field imaging \cite{tasse2012}.
On the one hand, the variation of the beam pattern with the direction (and with polarization) must be taken into account and, on the other hand, especially for continent-sized arrays, the curvature of Earth is no longer negligible and the projected baseline will have a third component, $w$. These two constraints destroy the simple FT relationship of Eq. \ref{eq:visibility} and Eq. \ref{eq:imaging}. These can be addressed using respectively the \textit{A-projection} \cite{tasse13} and \textit{W-projection} \cite{cornwell08} (or more recently \textit{W-Stacking} \cite{wsclean}).

Time and frequency integration can improve the quality and Signal-To-Noise (SNR) ratio of the image to a certain extent but an upper limit exists when smearing effects kill the correlation starting from the largest baselines. Long integrated images then require averaging images produced in smaller time windows (and smaller bandwidths) which provide enough SNR to perform deconvolution. Conversely, fast imaging will be limited by the amount of available data and therefore, by the noise level of the measured u-v components. However, a fast imaging mode is required when observing transient radio sources, therefore, the capability of detecting those transients will be limited by the noise level in the snapshot images. The accuracy, fidelity and SNR of the images and the required timescale to produce them put a limit on the minimum transient detection timescale.

There is scope for improving the image reconstruction algorithms by using recent results produced in the image/signal processing field. The application of the methods presented in Section \ref{sparse} represents an important milestone in the field of radio interferometric imaging by aperture synthesis.

\section{Principles of sparse image reconstruction}
\label{sparse}
	\subsection{Sparse representation}
	\label{sparserep}
		\subsubsection{Definition}

A discrete signal $x$ can be represented by Eq. \ref{eq:dict}.
\begin{equation}
x=\sum_{i=0}^{N} a_i \phi_i
\label{eq:dict}
\end{equation}
where the $a_i$ are the coefficients of $x$ in some space of representation $\Phi$, or \textit{dictionary}, composed of the \textit{atoms} $\phi_i$.
The signal $x$ is often expensive to store and there is interest of grasping the essential information about $x$ at a minimum cost.
One way to do this is to rely on sparse representations which enable data compression.
A signal is \textit{strictly} sparse when the family $\{a_i\}$ is almost null, e.g. most $\{a_i\}$ are null and $x$ can be represented with a small number of coefficients.
We define the support of $x$ as the set of non-null coefficients of $x$, and we can define the zero pseudo-norm $\ell_0$ as the cardinal of this support such as $||x||_{\ell_0}=\mathrm{card}(\mathrm{supp}(x))$.
Most natural signals are not strictly sparse but are ``compressible'' (or \textit{weakly} sparse), i.e. the essential information is gathered in some coefficients and the variation of the coefficient amplitude ranked by decreasing order converges quickly to 0 according to a power law \cite{starckbb}.
By defining a threshold on the coefficient amplitude, we can select only the most representative coefficients constituting $x$. Therefore $x$ can be approximated efficiently by its sparse representation $\tilde{x}$.

		\subsubsection{Dictionaries}

Depending on the nature of the signal, it is possible to find (and even build) a dictionary $\Phi$ in which the signal $x$ will have a relatively sparse representation. A dictionary is an indexed collection of atoms $\{\phi_i\}$, where the index can represent, a position (Dirac), a frequency (Fourier) or a scale (e.g. Wavelets). A dictionary can be represented as a matrix where the columns are the atoms.
	The choice of the correct dictionary is critical and for a given signal, the ``best'' dictionary is the one which provides the sparsest description. 
For example, a time series signal made of the sum of three cosine waves of different frequencies will not have a sparse representation in the time domain as a lot of coefficients are required to represent the signal with enough fidelity at each time step. However, in the Fourier domain, only three coefficients, corresponding to the three frequencies (knowing the parity of cosine function) are necessary to describe the signal entirely.	
Furthermore, a 2D image of the sky which only contains 1-pixel point sources will be sparse in the image domain (using Dirac functions to locate the non-null coefficient) but will not have a sparse representation in the 2D Fourier domain (where the information of the position is stored in the phase).
	
For an updated review on sparse representations and dictionaries, refer to \cite{starckbb,mallatbook}.
						
	\subsection{Solving inverse problems with convex optimization and proximal methods}
	\label{inverseproblem}

	\subsubsection{Inverse problems as convex optimization problems}
Sparsity and sparse representations could develop throughout the fields of signal/image processing thanks to the emergence of a new sampling theorem, \textit{compressed sensing} \cite{candes06,donoho06, candes08} which offers an alternate to the Shannon sampling theorem.
This new framework is extremely useful for formulating and solving ill-posed inverse problems.
A linear inverse problem can simply be formulated with Eq. \ref{eq:dic}.
\begin{equation}
y = \varphi(x) + n
\label{eq:dic}
\end{equation}
$y$ are the data, $x$ is the unknown signal to recover, $n$ is the noise and $\varphi$, the characteristic transform of the problem linking the wanted signal $x$ to the available data $y$. $\varphi$ operates as a degradation operator on $x$ and appears in various kind of problems as defined in \cite{starckbb}:
	\begin{itemize}
	\item Deconvolution problem: where $\varphi$ operates a convolution on $x$ filtering away the high spatial frequencies.
	\item Super resolution: $\varphi$ is also a convolution plus a subsampling operation.
	\item Inpainting: $\varphi$ is a binary operator which masks a substantial amount of information of $x$. This is important for the interferometric imaging problem where only discrete spatial frequencies are sampled. 
	\item Compressed Sensing decoding: $\varphi$ is an operator of random linear measurements of $x$ assuming that $x$ has a sparse representation in some dictionary $\Phi$.
	\end{itemize}

When these inversion problems are ill-posed (i.e. we are not sure of the existence and of the unicity of the solution, nor if the solution is stable under change of initial conditions) and when a good solution for $x$ is searched for, some prior knowledge on the structure of $x$ is required to reduce the space of acceptable solutions. Classical regularization techniques can be applied (such as the Tikhonov regularization \cite{tikhonov}) but sparse regularization techniques offer a wider range of methods. In the compressed sensing framework, three main conditions have to be met: i) the problem must be ill-posed, ii) a sparsifying dictionary must exist for $x$, iii) columns of $\varphi$ and $\Phi$ must be incoherent (the incoherence property is defined in \cite{candes08}).

The classical method for solving this problem is to find a solution $\tilde{x}$ that makes $\varphi(\tilde{x})$ close to the data $y$. For this, we define a misfit function as $||y-\varphi(x)||^2_{\ell_2}$ (where $\ell_2$ is the Euclidian norm) for which we want to find a minimum. We can formulate the minimization problem as Eq. \ref{eq:minsimple}
\begin{equation}
\min_x ||y-\varphi(x)||^2_{\ell_2}.
\label{eq:minsimple}
\end{equation}

We introduce a new quantity $\mathcal{R}(x)$ that will put a penalty on unwanted solutions. The minimization problem then becomes Eq. \ref{eq:minsimplereg}.
\begin{equation}
\min_x ||y-\varphi(x)||^2_{\ell_2} + \lambda\mathcal{R}(x).
\label{eq:minsimplereg}
\end{equation}

Let's assume that the signal $x$ is k-sparse in a dictionary $\Phi$ with coefficients $\{\alpha_i\}=\alpha$. $x$ can be written as in Eq. \ref{eq:dic}, $x=\Phi \alpha$ and its $\ell_0$ pseudo-norm is $||\alpha||_{\ell_0}=k$. In the context of inverse problems, we can transform Eq. \ref{eq:minsimple} into Eq. \ref{eq:minsparse}.
\begin{equation}
\begin{aligned}
\min_{\alpha} ||\alpha||_{\ell_0}  & &  \text{   s. t.   }  & & ||y-\varphi( \Phi \alpha) ||_{\ell_2} \leq \epsilon \\
\end{aligned}
\label{eq:minsparse}
\end{equation}

However, Eq. \ref{eq:minsparse} is non convex and falls to a NP-Hard combination problem, because of the $\ell_0$ pseudo-norm. We therefore introduce a convex relaxation by replacing the $\ell_0$ pseudo-norm with the $\ell_1$ norm (where $||\alpha||_{\ell_1}=\sum_i | \alpha |$).
Eq. \ref{eq:minsparse} reformulates in Eq. \ref{eq:minsparsel1} which, under it Lagrangian form, expressed as Eq. \ref{eq:minsparselag}.
\begin{equation}
\begin{aligned}
\min_{\alpha} ||\alpha||_{\ell_1}& &  \text{   s. t.   }  & & ||y-\varphi( \Phi \alpha) ||_{\ell_2} \leq \epsilon
\end{aligned}
\label{eq:minsparsel1}
\end{equation}
\begin{equation}
\min_{\alpha} \lambda ||\alpha||_{\ell_1} + \frac{1}{2}||y-\varphi( \Phi \alpha) ||^2_{\ell_2} \leq \epsilon
\label{eq:minsparselag}
\end{equation}
Which is of the same form as in Eq. \ref{eq:minsimplereg}. Here the regularization term $\mathcal{R}$ will enforce the sparsity of the coefficients $\alpha_i$.

\subsubsection{Proximal methods to solve convex problems}

The previous problem, because of the presence of the $\ell_1$ norm, is a convex but non-differentiable problem. Therefore, we can not apply smooth optimization techniques. In the scope of the convex analysis, the proximal theory introduced by Moreau in \cite{moreau1962} \footnote{Original paper which launched proximity operators.} generalizes the notion of projection to a specific class of convex, lower semicontinuous functions for which we can find a minimum. Interested readers should refer to \cite{starckbb,parikh2014proximal} for an introduction on proximal theory.

Problem \ref{eq:minsparselag} is of the form of Eq. \ref{eq:genericprox}. 
\begin{equation}
\min_\alpha \mathcal{H}(\alpha) = \mathcal{F}(\alpha) + \mathcal{G} (\alpha)
\label{eq:genericprox}
\end{equation}

where $\mathcal{F}$ and $\mathcal{G}$ are not necessarily differentiable, not infinite everywhere and their domains have non-empty intersection \cite{starckbb}. 

$\mathcal{F}$, as $\mathcal{R}$, is a penalty put on the sparsity of the solution and 
$\mathcal{G}$ is the data attachment term, or the data fidelity term, which promotes consistency of the solution with the input data.
The use of the \textit{proximity} operator helps us to determine a minimizer for Eq. \ref{eq:genericprox}.
The computation of a proximity operator for $\mathcal{H}$ is hard and splitting methods can be used to evaluate the proximity operators of $\mathcal{F}$ and $\mathcal{G}$ separately. Those methods split into three main classes \cite{eckstein,combettes2009} among which one can find the Forward-Backward splitting method which generalizes the classical gradient projection method as a constrained convex optimization (\cite{tseng1991,tseng2000} and reference therein).

If $\mathcal{G}$ is differentiable (with a L-Lipschitz continuous gradient) and $\mathcal{F}$ admits a \textit{proximal} operator, then we can solve the problem with the Forward-Backward (FB) splitting method using the iterative method depicted in Eq. \ref{eq:FBsplitting}.

\begin{equation}
x_{n+1}=\textrm{prox}_{\gamma_n \mathcal{F}}(x_n-\gamma_n \nabla \mathcal{G}(x_n))
\label{eq:FBsplitting}
\end{equation}
where $0 < \inf_{n\gamma_n} \leq \sup_{n\gamma_n} < 2/L$.
In Eq. \ref{eq:FBsplitting}, $\mathcal{F}=\lambda ||\alpha||_{\ell_1}$ and $\mathcal{G}=\frac{1}{2} ||y-\varphi( \Phi \alpha) ||^2_{\ell_2}$.


\section{Applying sparse reconstruction to radio interferometry}
\label{application}
	\subsection{Recasting the imaging inverse problem}
\label{recasting}

\begin{figure}[ptb]
\centering
\includegraphics[width=1.\textwidth]{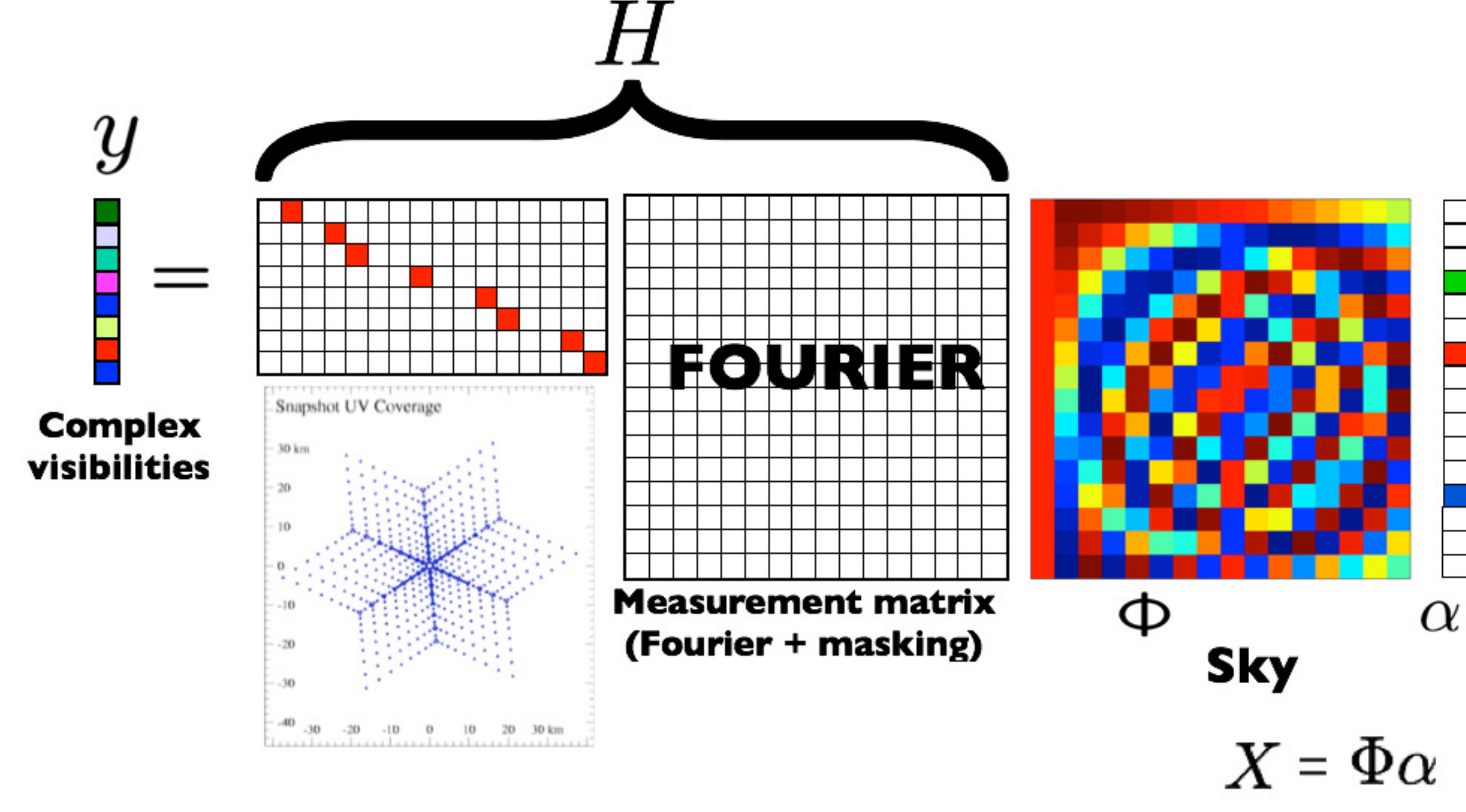}
\caption{Schematics of the inverse problem of imaging with $y$ the visibility data, $H$ the transform, combining the Fourier transform and the masking operator to account for the unmeasured part of the (u,v) plane, and $x$ which has a sparse representation of coefficients vector $\alpha$ in the dictionary $\Phi$.}
\label{radioimaging}
\end{figure}

The link between imaging by aperture synthesis in radio and the use of convex optimization was already made in the literature in \cite{wiaux09,Wenger2010,Li2011,carrillo12, garsden2015}.
The problem of radio interferometric imaging is an inverse problem as stated by Eq. \ref{eq:imaging} and as shown in Fig. \ref{radioimaging}.

In the context of the previous optimization framework, even the CLEAN algorithm can be interpreted as a matching-pursuit algorithm \cite{lannes97} using a dictionary with a single atom, the PSF.
In the specific context of LOFAR (and SKA) and low frequency imaging, we proposed an implementation of this problem in the LOFAR imagers (\textit{AWimager} \cite{tasse13}) under the name Sparse Aperture Synthesis Image Reconstruction (SASIR) \cite{garsden2015}. With SASIR integrated in the imager, it becomes straightforward to take into account the data format (Measurement Sets, the standard format of CASA software package \cite{NRAOcasa}) containing simulated or real data, as well as introducing direction dependent corrections such as the effect of the antenna beam (\textit{A-projection}) and of the non-coplanarity of the interferometer during wide-field imaging (\textit{W-projection}).

Theoretical introduction and numerical results on simulated and real LOFAR data are presented in \cite{garsden2015}\footnote{the source code is available at www.cosmostat.org}.
Among the different approaches, we can cite MORESANE \cite{moresane} which uses a greedy approach and smart image modeling using the Starlet transform \cite{starletbook,starckbb} and PURIFY \cite{Carrillo2014} which also use sparse representation and convex optimization.
Comparative tests are underway among the community of radio astronomers using the compressed sensing theory and a common testing framework, R.O.D.R.I.G.U.E.S.\footnote{accessible at http://rodrigues.meqtrees.net/scheduler/} \cite{rodrigues}, is being developed in the scope of the SKA precursors and will be able to furnish the tools to compare in parallel the performance of different imagers using the same input data and inspecting the same figure of merit on the reconstructions.
Hereafter, we describe the parameters and numerical results using SASIR only.
	
		\subsubsection{Choosing a dictionary}
Wavelet decomposition has impacted signal/image processing and data compression in a profound way.
Many different wavelets exist depending on the field and on the morphology of the signal (refer to \cite{mallatbook} for an exhaustive overview of wavelets). For astronomy, the \textit{starlet} transform (or Undecimated Isotropic Wavelet Transform -- IUWT) has demonstrated to provide efficient decompositions of extended complex sources while being computationally cheap.
This transform operates as a spatial filter on the image. The number of decomposition scales must be adapted to the complexity and to the variety of spatial frequencies in the data.
Some astrophysical structures present filamentary features which can be better represented with curvelets \cite{curvelets}. An ideal dictionary would be a joint set of dictionaries which accounts for all the different kind of shapes in the sky. Such representation however requires advanced signal reconstruction methods, such as MCA or GMCA \cite{bobin07}.

		\subsubsection{Choosing a minimization algorithm}
Our implementation of Eq. \ref{eq:FBsplitting} is based on the (Fast) Iterative Shrinkage Thresholding Algorithm -- (F)ISTA \cite{BeckFista09}.
With the implementation of Eq. \ref{eq:imaging}, the forward step computes the gradient and the backward step evaluates the proximal operator of $||.||_{\ell_1}$ which reduces to a soft-thresholding step on the wavelets coefficients of $x$.
Its implementation is simple and follows the algorithm \ref{Fista} (see \cite{garsden2015} for details).
Improvements of this algorithm include the implementation of a ``reweighting'' $\ell_1$ \cite{candesR1} which reduces the bias of the solution.
 \begin{algorithm}[!htb]
\caption{FISTA implementation}
\label{Fista}
\begin{algorithmic}[1]
\REQUIRE $\quad$ \\
A dictionary $\bm{\Phi}$.\\
The implicit measurement/degradation operator $\mathbf{A}$.\\
The original visibility data $\bm{V}$.\\
A detection level k.\\ 
The soft-thresholding operator:\\
$\text{SoftThresh}_{\lambda}(\bm{\alpha})=\prox_{\bm{\lambda}}(\bm{\alpha})=\left(\left(1-\frac{\lambda_j}{|\alpha_j|}\right)_{+} \alpha_j \right)_{1\leq j\leq N}$
        
\bigskip

\STATE {Initialize $\bm{\alpha}^{(0)}$=$\bm{0}$,  $t^{(0)}$=1, $\mathbf{x}^{(0)}$=$\bm{0}$ }

\FOR{$n=0$ to $N_{\max}-1$} 
 \STATE {$\bm{\beta}^{(n+1)} = \mathbf{x}^{(n)} + \mu \bm{\Phi}^{\text{T}} \mathbf{A}^{\text{T}}(\bm{V}-\mathbf{A}\bm{\Phi}\bm{\alpha}^{(n)})$}
 
 \STATE{$\mathbf{x}^{(n+1)}=\text{SoftThresh}_{\mu \bm{\lambda}_{j}} \bm{\beta}^{(n+1)}$}
 
 
  \STATE {$t^{(n+1)}=(1+\sqrt{1+4 (t^{(n)})^2})/2$}
    \STATE {$\gamma^{(n+1)}=(t^{(n)}-1)/t^{(n+1)}$}
        \STATE {$\bm{\alpha}^{(n+1)}=\mathbf{x}^{(n+1)}+\gamma^{(n+1)}(\mathbf{x}^{(n+1)}-\mathbf{x}^{(n)})$}
 \ENDFOR
 
 \STATE {\bf Return:} image $\bm{\Phi\bm{\alpha}}^{(N_{\max})}$.
\end{algorithmic}
\end{algorithm}

		\subsubsection{Choosing parameters}
In addition to the classical imaging parameters (number of pixels, pixel size on the sky, weighting scheme, ...), the gain and threshold definition has been updated in the context of convex optimization.
		The parameters $\mu$ in Algorithm \ref{Fista} is the relaxation parameter which impacts the convergence. $\mu$ taken to 1.0 is generally satisfying but it should be reduced to slow down the convergence on complex data. 
		The threshold parameter $\lambda$ is associated with the noise in the data and therefore is involved in the thresholding step of the wavelet coefficient. An improved thresholding scheme was adopted in SASIR and uses a different threshold $\lambda_i$ for each wavelet scale $i$. In each scale $i$ of the reconstruction residuals $R$, we compute $\sigma_i$ using the Median Absolute Deviation (MAD) as a robust noise estimator and taking $\lambda_i = n \sigma_i= n \frac{MAD(R_i)}{0.6745}$ \cite{madref}, $n$ embodying a detection level in this context.

	\subsection{Numerical tests}
	\label{numericaltests}
	In \cite{garsden2015}, we performed a series of tests to validate the quality of the image reconstruction. Astronomical images can be used for science if the photometry and the astrometry are accurate. We tested the reconstruction using two single point sources (\S\ref{angres}), a grid of point sources in a wide-field (\S\ref{photometry}) and testing on simulated and real LOFAR data containing extended radio emissions (\S\ref{extended}).
		\subsubsection{Angular resolution}
\label{angres}
We generated different LOFAR datasets containing two 1-Jy point sources at various angular distance ranging from 30'' to 30'. We jointly used the Cotton-Schwab CLEAN \cite{schwab84} to produce CLEAN images and SASIR, both inside \textit{AWimager} to produce the images. Results presented in \cite{garsden2015} demonstrated the capability of SASIR to image with super resolution giving a factor of $\approx$2 improvement as compared to the size of the CLEAN beam, on the same data and same weighting scheme. To beat the $\lambda/D$ limitation imposed by the sampling (and weighting) of the data in the (u,v) plane, the inpainting algorithm could restore unmeasured high spatial frequencies in the image. The maximum achievable resolution is therefore dependent on the size of the Fourier support, which is defined by the pixel size on the sky and the number of pixels in this support. We also compared the recovered flux density of the point sources, their astrometric errors in the presence of different level of noise (from a SNR of 3 up to 2000). Super-resolution is noticeable in a high SNR regime but the effective angular resolution falls back to that of CLEAN in the low SNR regime.

	\subsubsection{Photometry}
	\label{photometry}
\begin{figure}[tbp] 
\centering
\includegraphics[width=0.75\textwidth]{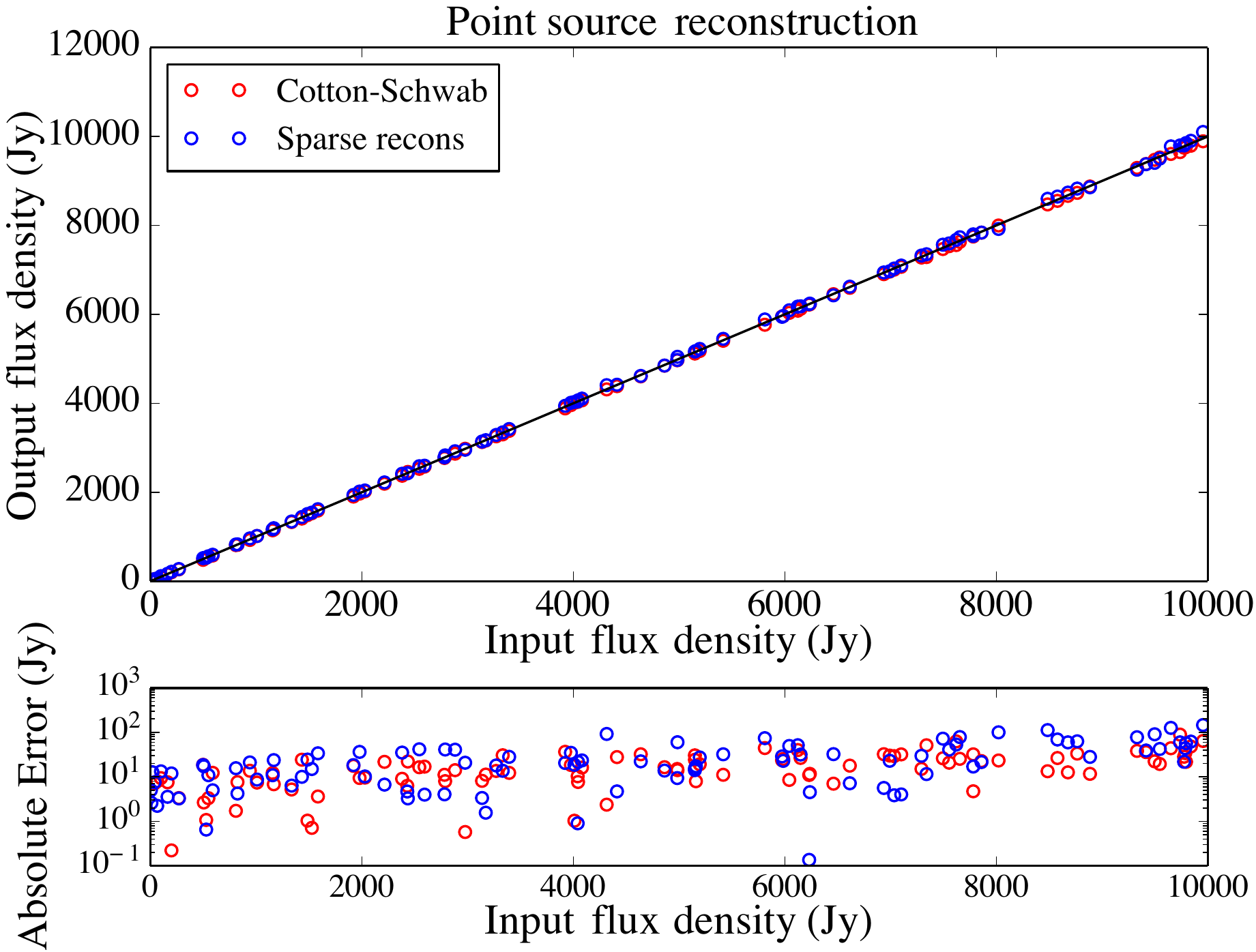}
\caption{(Top) Total flux density reconstruction for a set of point sources with original flux density spanning over 4 orders of magnitude with the original flux density (x-axis) vs. the recovered flux density (y-axis). (Bottom) scatter plot of the absolute error for each source. The recovered flux densities for Cotton Schwab CLEAN (red) and SASIR (blue) are represented on a linear scale, whereas the absolute error is on a logarithmic scale for clarity. Perfect reconstruction lies along the black line. Adapted from \cite{garsden2015}.}
\label{fig:grid}
\end{figure}

We created a sky model containing a rectangular grid of 10$\times$10 point sources, distributed evenly in a field of $8^o\times8^o$ with flux density values ranging from 1 to 10$^4$ Jy. After the image reconstruction, we used \textit{PyBDSM}\footnote{See http://www.lofar.org/wiki/doku.php for more information.} as a source finder to locate and match the sources between the CLEAN and the Sparse reconstruction, as well as with the original sky model. The source finder could also give a measure of the flux densities of the detected source for each method. Fig. \ref{fig:grid} shows the output flux densities versus the input flux densities of the sky model for CLEAN (in red) and SASIR (in blue). The photometry using SASIR is comparable to that of CLEAN. The latter being particularly well fitted for point sources. We expect no major improvement of the flux density with SASIR. Nonetheless, the difference can still be drawn while imaging extended emission data.

	\subsubsection{Extended emission}
	\label{extended}
\begin{figure}[tbp] 
\centering
\includegraphics[width=1.\textwidth]{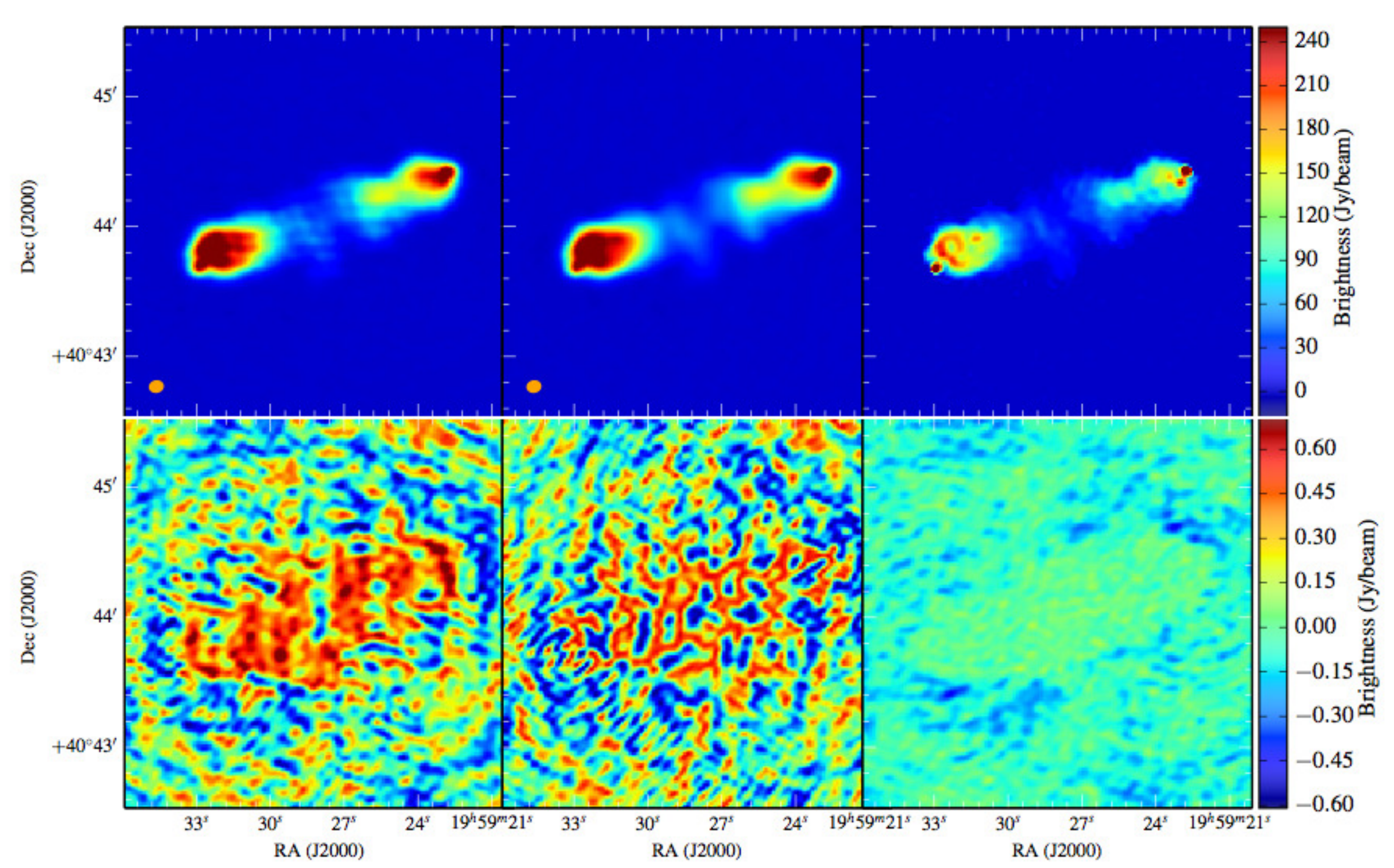}
\caption{Reconstructed images of Cygnus A from a real LOFAR observation using Cotton Schwab CLEAN (left column), MS-CLEAN (middle column), and the sparse reconstruction with SASIR (right column). First row are the restored images and second row are the residual images. The sparse reconstruction presents a higher angular resolution and a lower residual level than that obtained with the two other methods. Adapted from \cite{garsden2015}.}
\label{fig:cygnusA}
\end{figure}

We simulated a Measurement Set containing a sky model of the W50 super nova remnant adapted from high frequency radio maps \cite{dubner98}. As the emission is extended, we produced images using Cotton-Schwab CLEAN and Multiscale CLEAN. The sparse reconstruction with SASIR enabled a reconstruction with higher effective angular resolution and a lower level of the reconstruction residuals (by a factor of $\approx$ 5).
As a final test, we used a calibrated LOFAR dataset containing real data taken on Cygnus A, a strong radio source presenting two bright radio lobes at low frequencies. The sparse reconstruction came with higher resolution and better reconstruction residuals as compared to Cotton-Schwab CLEAN and MS-CLEAN. We present on Fig. \ref{fig:cygnusA}, the reconstruction and residuals for the different methods.
In \cite{garsden2015}, the LOFAR map obtained at 150 MHz with SASIR was compared to radio contours obtained with the VLA at twice the frequency (325 MHz). The recovered super-resolved structures match real structures seen in the VLA radio map which shows that SASIR is able to recover missing spatial information with high fidelity.


\section{Conclusions and Future developments}
Sparse representation, convex optimization and proximal theory offer a new framework to develop advanced tools for radio imaging using aperture synthesis. Tests and developments of these new kind of radio imagers are on going and are developing in the context of LOFAR and SKA. We presented with SASIR, an example of solving an inverse problem through inpainting in the visibility plane.
Sparse reconstruction with SASIR is now being tested on various objects such protoplanetary disk imaging, planetary radio emissions, transients and most recently in the scope of radio weak lensing studies \cite{piresWL,patel2014} in the context of SKA \cite{patel2015}. In cosmology, blobs of dark matter are known to induce shearing on the image of galaxies and there is scope for evaluating the quality of the galaxy shape fitting on sparse reconstructed images and compare the results to that of CLEAN.
Future implementations of the sparse reconstruction will extend to a third dimension to focus on transient detection and spectral imaging.
Sparse methods provide robust and realistic images from sparsely sampled data. This sparsity can be used on purpose in the design of an array (e.g. \cite{Fannjiang2013}). Instruments such as the SKA will provide a tremendous amount of redundant data that are difficult to calibrate. By applying the new framework of \textit{Compressed Sensing} in the hardware part (by sensing randomly or using random antenna configurations), it could enable the implementation of fast imaging modes which partly alleviates the problem of data storage, which is particularly heavy when the data to store are raw visibilities. These new methods can be applied, on a longer term, to instrumental calibration directly in order to deduce relevant instrumental parameters from sparse data. The use of these methods is only at its beginning in (radio)astronomy, but its potential scientific benefits could represent a big leap forward in the field of astronomical data processing and data imaging.
\acknowledgments
We would like to thank the anonymous reviewers for their suggestions and comments. We acknowledge the financial support from the UnivEarthS Labex program of Sorbonne Paris Cit\'e (ANR-10-LABX-0023 and ANR-11-IDEX-0005-02), from the European Research Council grant SparseAstro (ERC-228261) and from the Physis project (H2020-LEIT-Space-COMPET).
LOFAR, the Low Frequency Array designed and constructed by ASTRON, has facilities in several countries, that are owned by various parties (each with their own funding sources), and that are collectively operated by the International LOFAR Telescope (ILT) foundation under a joint scientific policy. For more information on LOFAR, visit www.lofar.org.

\bibliographystyle{unsrt}
\bibliography{refs.bib}


\end{document}